\documentclass[smallextended]{svjour3}
\smartqed
\usepackage{graphicx}
\usepackage{amsmath}
\usepackage{amssymb}

\begin{document}
		
	\title{Quantum key distribution based on orthogonal state encoding}
	\author{Hao Shu}
	\institute{Hao Shu\at
		College of Mathematics, South China University of Technology,Guangzhou, 510641, P. R. China
		\\
		\email{Hao\_B\_Shu@163.com}
	}

	\date{}	
	
	\maketitle
	
	\begin{abstract}
		Quantum key distribution(QKD) is one of the most significant areas in quantum information theory. For nearly four decades, substantial QKD schemes are developed. In early years, the security of QKD protocols is depend on switching different bases, namely based on non-orthogonal state encoding. The most famous example is the BB84 protocol. Later, other techniques were developed for orthogonal state encoding. Examples of such protocols include the GV protocol and order-rearrangement protocols. In this paper, we present two QKD protocols based on orthogonal state encoding. One of them does not need to employ order-rearrangement techniques while the other needs to. We provide analyses for them, demonstrating that they are highly efficient when considering consumptions of both qubits and classical bits. Furthermore, the employment of maximally entangled states could be less than previous ones and so the measurement efficiency could be increased. We also modify the protocols for implementing in noisy channels by applying the testing state method.
		
	
	\end{abstract}
	
	\section{Introduction}
	
	In information theory, cryptography is always one of the most important fields. Unfortunately, the most useful cryptosystem nowadays, the RSA system, is not secure in quantum era\cite{S1994Algorithms}, for the essential reason that its security depends on the low capacity of classical computations. To obtain the unconditional security, cryptographic schemes with the security only depends on physical laws are needed. The only classical cryptosystem proven to be secure is encoding messages with an one-time pad. However, transmitting an one-time pad by classical channels could be totally insecure since classical messages can be cloned without being detected. On the other hand, quantum effects can provide possibilities for transmitting an one-time pad with the security only depends on physical laws and can be proven mathematically. Such a task is called quantum key distribution (QKD).
	
	The first QKD protocol was proposed in 1984\cite{BB1984Quantum}, obtaining the security by switching two mutually unbiased bases, whose security has been proven\cite{SP2000Simple}. After the BB84 protocol, several BB84-like protocols were proposed, such as Ekert's protocol\cite{E1991Quantum}, BBM92 protocol\cite{BB1992Quantum}, six-state protocol\cite{B1998Optimal} and others\cite{SP2000Simple,CB2002Security}. In nearly four decades, especially in recent years, substantial variants of QKD protocol such as device-independent (DI) QKD\cite{LZ2022Toward,ZL2022A,XL2021Overcoming}, measurement-device independent (MDI) QKD\cite{YC2016Measurement,FL2021Measurement,XL2022Breaking}, Twin-field QKD\cite{LY2018Overcoming,WY2022Twin}, continuous variable (CV) QKD\cite{LL2021Homodyne,LZ2022Automated}, coherent-one-way QKD\cite{WB2014A,GX2022Simple} as well as QKD over a network\cite{FL2022Robust} have been developed both theoretically and experimentally. Other methods that can be employed to transmit private message securely were also discovered, for example, quantum secure direct communication (QSDC)\cite{LL2002Theoretically,ZD2017Quantum,DL2003Two,BF2002Deterministic,BE2001Secure,WL2022Quantum,PL2020Experimental,WZ2006Quantum,DL2007Quantum} and teleportation\cite{BB1993Teleporting}. However, QKD is still the most-used one since it might be more reliable in a lossy channel and easier to be implemented.
	
	Most QKD protocols obtain the security by employing non-orthogonal states. It was not until 1995 that the first cryptographic protocol based on orthogonal states was published\cite{GV1995Quantum}. The idea of the protocol is sending states with a time delay such that eavesdroppers can never get an entire state without being detected. Another technique can be employed to implement a QKD protocol based on orthogonal states is the order-rearrangement. Protocols employing this technique can be found in \cite{DL2003Controlled,SA2014Protocols,ZX2006Secure,YS2014Two,SP2013Beyond}. Other protocols include \cite{GL2001Quantum,AB2014Orthogonal,H2011Quantum,SB2016Secure,N2009Counterfactual,AB2010Experimental,AS2013Semi}.
	
	Although non-orthogonal state encoding seems already mature and enough, orthogonal state encoding might provide extra advantages such as saving quantum operations. Even if considering theoretical meaning only, knowing how orthogonal state can be employed for coding has its own interests, since orthogonal states can be distinguished without errors.
	
	Besides designing QKD protocols, there is another problem. In practise, Channels employed to implement a QKD protocol are always noisy. Therefore, the robustness of a protocol over noisy channels has to be investigated. Previous works include \cite{LZ2009Fault,LD2008Efficient,SD2010Efficient,BG2004Robust} for collective noises, \cite{SS2007Degenerate,FW2008Lower,FM2001Enhanced,CR2011Experimental} for Pauli noises and \cite{TP2015Applications,SS2015Controlled,OS2013Dissipative,TM2000Decoherence,XY2016Protecting} for amplitude damping(AD) and phase damping(PD) noises. There are also other researches on noises, for example \cite{ST2015Which,SB2008Squeezed,SO2012The,TB2015Quasiprobability,TB2016Tomograms,SZ2021Entanglement}.
	
	In this paper, two QKD protocols based on orthogonal state encoding are proposed in section II while efficiency analyses are given in section III, with a comparison to several previous protocols. The discussions of the security are given in section IV and implementing over noisy environments would be argued in section V. The last section, section VI, is devoted to conclusions. The protocols employ different technologies to make orthogonal state encoding effective. In details, Protocol I employs the non-locality of states while protocol II employs the order-rearrangement technology. Comparing with certain previous ones, our protocols are highly efficient when considering consumptions of both qubits and classical bits. Furthermore, the employment of maximally entangled states could be less and so the measurement efficiency could be increased. And on the other hand, they can be robust over several noisy channels, namely they could be implemented in noisy channels as in noiseless ones without errors after modifications.

	\section{Two protocols}
	
	The protocols are stated as follow, in which the subscripts A, B denote the partite and we will employ $|a+b\rangle$ to represent $|a\rangle+|b\rangle$ for simplifying the notation. Here, a partita of a state means its correspondent particle in the local subsystem.
    \\

	\textbf{Protocol I:}
	\\
	
	\textbf{Step 1}: Alice and Bob agree to encode $00, 11, 01, 10$ by states $|00\rangle_{AB}, |11\rangle_{AB},\\ |\varphi\rangle_{AB}=\frac{1}{\sqrt{2}}|01-10\rangle_{AB}, |\varphi'\rangle_{AB}=\frac{1}{\sqrt{2}}|01+10\rangle_{AB}$, respectively, in $C^{2}\otimes C^{2}$.
	
	\textbf{Step 2}: To share a N-2-bit key string (namely, a bit string whose length is N with every position has 2 bits information, since the coding states are now in $C^{4}$), Alice creates a string of 2N states chosen randomly in $S=\left\{ |00\rangle_{AB}, |11\rangle_{AB}, |\varphi\rangle_{AB}, |\varphi'\rangle_{AB} \right\}$, which is only known by her.
	
	\textbf{Step 3}: Alice creates $\frac{N}{2}$ decoy states, all be $|+\rangle_{B}=\frac{1}{\sqrt{2}}|0+1\rangle_{B}$ and inserts them into the state string in step 2 randomly. Now Alice has a string with 2.5N states and she records the positions of decoy states by a 2.5N-bit string $r=r_{1}r_{2},...r_{2.5N}$. In more details, $r_{i}=1$ if the i-th state is a decoy state and $r_{i}=0$, otherwise.
	
    \textbf{Step 4}: Alice sends the partita B of the state string to Bob.
	
	\textbf{Step 5}: After receiving the particles, Bob publicly announces this fact.
	
	\textbf{Step 6}: After Alice receives Bob's receipt, she sends the partita A of the state string together with the string r to Bob.
	
	\textbf{Step 7}: Bob receives the state string and the string r. He then measures decoy states via basis $\left\{ |+\rangle=\frac{1}{\sqrt{2}}|0+1\rangle, |-\rangle=\frac{1}{\sqrt{2}}|0-1\rangle \right\}$, and other states via basis $S=\left\{ |00\rangle_{AB}, |11\rangle_{AB}, |\varphi\rangle_{AB}, |\varphi'\rangle_{AB} \right\}$.
	
	\textbf{Step 8}: Alice and Bob run the checking procedure as follow. Bob publishes all his outcomes on decoy states and the outcomes on half of other states (let us call them checking states) chosen randomly. Bob publishes his outcomes on checking states together with the positions of them. Alice verifies whether the checking states are agreed with what she created while Bob verifies whether the decoy states are $|+\rangle$. They calculate the error rates.
	
	\textbf{Step 9}: If the error rates are acceptable on both decoy states and checking states, Alice and Bob agree a secret key by the outcomes of the remaining N states, which are neither decoy states nor checking states.
	
	\textbf{Step 10}: Alice and Bob repeat the above procedure until they share a sufficiently long secret key and run error correcting and privacy amplification procedures if needed.
\\

    \textbf{Protocol II:}
\\

    \textbf{Step 1}: Alice and Bob agree to encode $00, 11, 01, 10$ by states $|00\rangle_{AB}, |11\rangle_{AB},\\ |\varphi\rangle_{AB}=\frac{1}{\sqrt{2}}|01-10\rangle_{AB}, |\varphi'\rangle_{AB}=\frac{1}{\sqrt{2}}|01+10\rangle_{AB}$, respectively, in $C^{2}\otimes C^{2}$.

    \textbf{Step 2}: To share a N-2-bit key string, Alice creates a string of 2N states chosen randomly in $S=\left\{ |00\rangle_{AB}, |11\rangle_{AB}, |\varphi\rangle_{AB}, |\varphi'\rangle_{AB} \right\}$ which is only known by her.

    \textbf{Step 3}: Alice divides the 2N states into N blocks such that each block includes two adjacent states. Alice chooses a random N-bit string $s=s_{1},s_{2},...,s_{N}$ and exchanges the order of the B partita of the first state and the A partita of the second state in the i-th block if $s_{i}=1$.

    \textbf{Step 4}: Alice sends all partite of the state string to Bob.

    \textbf{Step 5}: After receiving the particles, Bob publicly announces this fact.

    \textbf{Step 6}: After Alice receives Bob's receipt, she sends the string s to Bob.

    \textbf{Step 7}: Now, Bob has the state string and the string s. He then reorders the states by information in the string s, recovering them and then measures via basis S.

    \textbf{Step 8}: Alice and Bob run a checking procedure as follow. Bob publishes the positions and outcomes on half of states (let us call them checking states) chosen randomly. Alice verifies whether the checking states are agreed with what she created and calculates the error rate.

    \textbf{Step 9}: If the error rate is acceptable, Alice and Bob agree a secret key by the outcomes of the remaining N states, which are not employed as checking states.

    \textbf{Step 10}: Alice and Bob repeat the above procedure until they share a sufficiently long secret key and run error correcting and privacy amplification procedures if needed.
	\\
	
	Some issues should be illustrated. Firstly, protocol I is not totally based on orthogonal states, since the decoy states are not orthogonal to S. But at least the coding states are orthogonal and thus can be viewed as orthogonal state encoding. Protocol II is totally based on orthogonal states, as all states in it are orthogonal. Secondly, although inserting decoy states in protocol I can be done by an order-rearrangement, it can also be done without that, just preparing states with the chosen order recorded by r. Thirdly, in above protocols, we assume that the legitimate partner can employ quantum memories. However, in protocol II, this assumption can be removed by Bob guesses $s_{i}$ for every block and measures the receiving states immediately depending on the guessing order. In details, Bob reorders the state in the $i$-th block and measures them if he guesses $s_{i}=1$, while he measures the states directly if he guesses $s_{i}=0$. Here, the bits in a block that Bob wrongly guesses the order need to be aborted. The procedure is in the same position as basis sifting. Fourthly, One might doubt that how Bob knows that he receives the state without measuring it immediately. In fact, on one hand, deciding whether a signal is received not necessarily needs to measure the state. On the other hand, Bob does not need to know exactly whether the states are received. The main point is that Alice is required to implement step 6 after Bob storages the signals sent by Alice in step 4. The signals can be empty (in such case, Bob storages nothing) or not. If some of the signals are empty, then the corresponding joint measurements will fail, since a joint measurement can be effective only if two detectors provide outcomes. On the other hand, a joint measurement with one signal contains nothing while the other contains two particles should provide random outcome and thus can not make the eavesdropper benefit. Finally, how many bits are needed to be employed for channel estimating depends on key analysis and might be less than half. We choose half of coding states for estimating here because we want to compare the protocols to ordinary BB84 protocol.
	
	\section{Analyses of the protocols}
	
	Before discussing the security, let us give analyses of the protocols. We will provide analyses of the protocols on both consumption in qubits and classical bits and demonstrate that our protocols could be more efficient than the previous ones. For a criterion of efficiency, one can use $e=\frac{c}{q+b}$, defined in \cite{C2000Quantum}, where c denotes the bits in the key, q denotes the consumption of qubits and b denotes the consumption of classical bits. However, we note that one might directly compare the consumption of qubits and classical bits separately with other protocols, since, as mentioned in \cite{C2000Quantum}, the setting is not realistic.
	
	\subsection{Consumption on both qubits and classical bits}
	
	In protocol I, to generate a N-2-bit key string, Alice and Bob consume 2N states in $C^{2}\otimes C^{2}$ and $\frac{N}{2}$ single qubit states. Totally, they consume 4.5N qubits. On the other hand, Alice and Bob consume 2.5N classical bits for publishing the string r, 2N classical bits for publishing the positions of checking states and 2N classical bits for publishing the measurement outcomes of N checking states. In more details, for example, Bob sends a string $b=b_{1}b_{2},...,b_{2N}$ to Alice with $b_{i}=0$ represents that the i-th state is not a checking state, and $b_{i}=1$, otherwise. And for the N checking states, Bob sends a string $c=c_{1}c_{2},...,c_{N}$ to Alice with $c_{j}=0,1,2,3$ represents that the measurement outcome of the j-th state is $|00\rangle, |11\rangle, |\varphi\rangle, |\varphi'\rangle$, respectively. Of course, they have to consume another three classical bits including Bob declares his receipt in step 5, Alice and Bob publish whether their error checking procedures are passed. The total classical bits needed (for classical communications) in such a protocol are nearly 6.5N. Equivalently, they consume nearly 2.25N qubits and 3.25N classical bits for a N-bit key string. Hence, $e=\frac{1}{5.5}$.
	
	In protocol II, to generate a N-2-bit key string, Alice and Bob consume 2N states in $C^{2}\otimes C^{2}$. Totally, they consume 4N qubits. On the other hand, Alice and Bob consume N classical bits for publishing the string s, 2N classical bits for publishing the positions of checking states and 2N classical bits for publishing the outcomes of N checking states. In more details, for example, Bob sends a string $d=d_{1}d_{2},...,d_{2N}$ to Alice with $d_{i}=0$ represents that the i-th state is not a checking state, and $d_{i}=1$, otherwise. And for the N checking states, Bob sends a string $e=e_{1}e_{2},...,e_{N}$ to Alice with $e_{j}=0,1,2,3$ represents that the outcome of the j-th state is $|00\rangle, |11\rangle, |\varphi\rangle, |\varphi'\rangle$, respectively. Of course, they have to consume another two classical bits including Bob declares his receipt in step 5, Alice publishes whether the error checking procedure is passed. The total classical bits (for classical communications) needed in such a protocol are nearly 5N. Equivalently, they consume nearly 2N qubits and 2.5N classical bits for a N-bit key string. Hence, $e=\frac{1}{4.5}$.
	
	\subsection{Comparing with previous protocols}
	
	Note that in the ordinary BB84 protocol\cite{BB1984Quantum}, to agree a N-bit key string, Alice and Bob have to consume 4N qubits, 4N classical bits for Bob inform Alice which bases he chose to measure, 4N classical bits for Alice inform Bob which states are discarded, 2N classical bits for Alice inform Bob which states are checking states and N classical bits for Alice inform Bob the outcomes on checking states. Totally, they consume nearly 4N qubits and 11N classical bits. Hence, $e=\frac{1}{15}$.
	
	In the modified BB84 protocol\cite{SP2000Simple} with a Hadamard gate, it consumes 2N qubits and 2N classical bits for publishing operations on states and 2N classical bits for publishing positions of checking states and another N classical bits for publishing the outcomes on checking states. Totally, it consumes 2N qubits and nearly 5N classical bits for a N-bit key string. Hence, $e=\frac{1}{7}$.

    In previous protocols based on order-rearrangement of orthogonal states\cite{DL2003Controlled,SA2014Protocols,ZX2006Secure,YS2014Two,SP2013Beyond}, the consumptions of qubits and classical bits are not less than protocol II above. For example, in \cite{DL2003Controlled}, the consumption of states is equal to protocol II but has to employ more entangled states and a full Bell measurement, whose efficiency is low experimentally. Protocol II does not need to measure a full set of Bell states, which might be more efficient in implementing. And on the other hand, the previous protocol requires a four-state rearrangement in four cases while protocol II only requires a two-state rearrangement in two cases, which could be more realistic.
	
	\section{Security}
	
	Let us focus on the security of the protocols. Assume that there is an eavesdropper, says Eye, who wants to steal the secret key of Alice and Bob. We would assume that Alice and Bob hold authenticated classical channels which might not be private, and quantum channels without any further assumption. Namely, Eve might eavesdrop the classical communications but with no abilities to forge messages or pretend to be one of the legitimated parties, while she can do anything under physical laws in quantum channels. The security is in the sense that Eve can not get enough information on the secret key or she would create errors which are detectable by the checking procedure. We also assume that Eve provides a collective attack, namely, Eve attacks each state sent by Alice with the same strategy.
	
	We discuss three kinds of attacks. Eve might intercept a state sent by Alice and implement one of the three actions. Firstly, she might add auxiliary partite (her partite) then do a transformation and resend the state to Bob , namely entangle the state with a probe (let us call this a purified attack). Secondly, she might take the state herself and send another state created by her to Bob, instead (let us call this a substituted attack). Thirdly, she might measure the state and resend it to Bob (let us call this a measure-resend attack). Let us also assume that in protocol I, Eve can implement different attacks in step 4 and step 6. Note that such attacks are not general and more general analysis might be provided later.
	
	\subsection{purified attack}
	
	Let us analyse purified attacks firstly. For such attacks, the decoy states in the protocol I (step 3) can be aborted, and so the efficiency can be increased.
	
	\subsubsection{Eve purifies via single qubits}
	
	If Eve purifies states sent by Alice via single qubits, the security of the protocols (both protocol I and protocol II) could correspond to previous ones, such as \cite{SB2016Secure} or \cite{DL2003Controlled}. For example, if Eve purifies states sent by Alice via basis
	$\left \{ |j\rangle |j=0,1 \right \}$, which would change states $|00\rangle, |11\rangle, |\varphi\rangle, |\varphi'\rangle$ into $|00\rangle_{AB}|00\rangle_{EE'}, |11\rangle_{AB}|11\rangle_{EE'},\\ |\varphi_{p}\rangle_{ABEE'}=\frac{1}{\sqrt{2}}|0101-1010\rangle_{ABEE'}, |\varphi_{p}'\rangle_{ABEE'}=\frac{1}{\sqrt{2}}|0101+1010\rangle_{ABEE'}$, respectively, where E and E' are partite of Eve, then Alice and Bob can detect Eve on checking states $|\varphi\rangle$ or $|\varphi'\rangle$. In more details, $|\varphi_{p}\rangle_{ABEE'}=\frac{1}{\sqrt{2}}|0101-1010\rangle_{ABEE'}=\frac{1}{\sqrt{2}}(|\varphi\rangle|\varphi'\rangle+|\varphi'\rangle|\varphi\rangle)_{ABEE'}$. When Bob measures the state via basis S on partite A and B, he will get an outcome $|\varphi\rangle$ or $|\varphi'\rangle$ with equal probabilities. The calculation of $|\varphi'\rangle$ is essentially same. The error rate caused by Eve and detectable by Alice and Bob is now $\frac{1}{4}$, for Alice and Bob choose an entangled state for checking with probability $\frac{1}{2}$ and get an error outcome with probability $\frac{1}{2}$, if so.
	
	If Eve purifies states via other bases, the arguments are similar. Note that if so, errors could occur on all checking states. For example, Eve purifies states via basis $\left\{ |+\rangle=\frac{1}{\sqrt{2}}|0+1\rangle, |-\rangle=\frac{1}{\sqrt{2}}|0-1\rangle \right\}$, then an error occurs with probability $\frac{1}{2}$ for each checking state and so the error rate is $\frac{1}{2}$.
	
	\subsubsection{Eve purifies via two-qubit states}
	
	Now assume that Eve attacks by purifying states via S. This attack is not suitable for protocol I since in protocol I, Eve can only obtain one partita of states in the same time, if she tries to employ such an attack. Let us analyse for protocol II.
	
	Since Alice changes the order of the two partite of the two states, Eve guesses the order correctly with probability $\frac{1}{2}$. If she guesses the order wrongly, she will purify the states wrongly and create errors, which can be detected with probabilities. In more details, let a block of two states be $|b\rangle_{xy}=|x\rangle_{12}|y\rangle_{34}$, where $|x\rangle$ and $|y\rangle$ are states in S. If Eve guesses the order of states wrongly, she will purify states via S on partite 1, 3, and partite 2, 4, respectively. $|b\rangle_{xy}$ can be one of

\begin{equation}
\begin{aligned}
	 b_{\varphi\varphi}=|\varphi\rangle_{12}|\varphi\rangle_{34}, b_{\varphi'\varphi'}=|\varphi'\rangle_{12}|\varphi'\rangle_{34},
	b_{00}=|00\rangle_{12}|00\rangle_{34}, b_{11}=|11\rangle_{12}|11\rangle_{34},
	\\
	b_{\varphi\varphi'}=|\varphi\rangle_{12}|\varphi'\rangle_{34}, b_{\varphi'\varphi}=|\varphi'\rangle_{12}|\varphi\rangle_{34},
	b_{\varphi0}=|\varphi\rangle_{12}|00\rangle_{34}, b_{0\varphi}=|00\rangle_{12}|\varphi\rangle_{34},
	\\
	b_{\varphi'0}=|\varphi'\rangle_{12}|00\rangle_{34}, b_{0\varphi'}=|00\rangle_{12}|\varphi'\rangle_{34},  b_{\varphi1}=|\varphi\rangle_{12}|11\rangle_{34}, b_{1\varphi}=|11\rangle_{12}|\varphi\rangle_{34},
	\\
	b_{\varphi'1}=|\varphi'\rangle_{12}|11\rangle_{34}, b_{1\varphi'}=|11\rangle_{12}|\varphi'\rangle_{34},
	b_{01}=|00\rangle_{12}|11\rangle_{34}, b_{10}=|11\rangle_{12}|00\rangle_{34}.\nonumber
\end{aligned}
\end{equation}

	 with equal probability. Let us calculate the error rate for each case. Assume that $|b\rangle_{xy}$ becomes $|b\rangle_{xyp}$ after being purified via S on partite 1, 3 and partite 2, 4, respectively. And let Eve's partite be $E_{1}, E_{2}, E_{3}, E_{4}$.

\begin{equation}
\begin{aligned}
	b_{\varphi'\varphi'}=&|\varphi'\rangle_{12}|\varphi'\rangle_{34}=\frac{1}{2}|0101+0110+1001+1010\rangle_{1234}
	\\
	=&\frac{1}{2}|0011+0110+1001+1100\rangle_{1324}=\frac{1}{2}|0011+\varphi'\varphi'-\varphi\varphi+1100\rangle_{1324}.
    \\
	  b_{\varphi'\varphi'p}=&\frac{1}{2}|00110011+\varphi'\varphi'\varphi'\varphi'-\varphi\varphi\varphi\varphi+11001100\rangle_{1324E_{1}E_{2}E_{3}E_{4}}
	  \\
	  =&\frac{1}{4}(|0011\rangle|\varphi'\varphi'-\varphi\varphi\rangle+|1100\rangle|\varphi'\varphi'-\varphi\varphi\rangle+|\varphi\varphi\rangle|0011+1100-\varphi'\varphi'-\varphi\varphi\rangle
	  \\
	  &+|\varphi'\varphi'\rangle|0011+1100+\varphi'\varphi'+\varphi\varphi\rangle+|\varphi\varphi'\rangle|0011-1100\rangle
	  \\
	  &+|\varphi'\varphi\rangle|0011-1100\rangle)_{1234E_{1}E_{2}E_{3}E_{4}}.\nonumber
\end{aligned}
\end{equation}

	  The probability of getting the correct outcome by measuring via S on partite 1, 2 is $\frac{3}{10}$ and so the error rate is $\frac{7}{10}$. Similarly, error rates for $b_{\varphi\varphi p}$, $b_{\varphi\varphi' p}$ and $b_{\varphi'\varphi p}$ are all $\frac{7}{10}$.
	
\begin{equation}
\begin{aligned}
	b_{\varphi0}=&|\varphi\rangle_{12}|00\rangle_{34}=\frac{1}{\sqrt{2}}|0100-1000\rangle_{1234}=\frac{1}{\sqrt{2}}|0010-1000\rangle_{1324}
	\\
	=&\frac{1}{2}|00\varphi'-00\varphi-\varphi'00+\varphi00\rangle_{1324}.
    \\
	 b_{\varphi0p}=&\frac{1}{2}|00\varphi'00\varphi'-00\varphi00\varphi-\varphi'00\varphi'00+\varphi00\varphi00\rangle_{1324E_{1}E_{2}E_{3}E_{4}}
	 \\
	 =&\frac{1}{4}(|00\rangle|\varphi+\varphi'\rangle|00\rangle|\varphi'-\varphi\rangle-|00\rangle|\varphi'-\varphi\rangle|\varphi'-\varphi\rangle|00\rangle
	 \\
	 &+|\varphi\rangle|00\rangle|00\varphi'+00\varphi+\varphi'00+\varphi00\rangle
	 \\
	 &+|\varphi'\rangle|00\rangle|00\varphi'+00\varphi-\varphi'00-\varphi00\rangle)_{1234E_{1}E_{2}E_{3}E_{4}}.\nonumber
\end{aligned}
\end{equation}

	 The probability of getting the correct outcome (that is $|\varphi\rangle$) by measuring via S on partite 1, 2 is $\frac{1}{4}$ and so the error rate is $\frac{3}{4}$. Similarly, error rates for $b_{\varphi'0p}$, $b_{\varphi'1p}$ and $b_{\varphi 1p}$ are all $\frac{3}{4}$. And on the other hand, the probability of getting the correct outcome by measuring via S on partite 3, 4 (that is $|00\rangle$) is $\frac{1}{2}$ and so the error rate is $\frac{1}{2}$. Similarly, error rates for $b_{0p\varphi'}$, $b_{1p\varphi'}$ and $b_{1p\varphi}$ are all $\frac{1}{2}$.

\begin{equation}
\begin{aligned}
	b_{01}=&|00\rangle_{12}|11\rangle_{34}=|0011\rangle_{1234}=|0101\rangle_{1324}=\frac{1}{2}|\varphi\varphi+\varphi\varphi'+\varphi'\varphi+\varphi'\varphi'\rangle_{1324}.
    \\
	  b_{01p}=&\frac{1}{2}|\varphi\varphi\varphi\varphi+\varphi\varphi'\varphi\varphi'+\varphi'\varphi\varphi'\varphi+\varphi'\varphi'\varphi'\varphi'\rangle_{1324E_{1}E_{2}E_{3}E_{4}}
	  \\
	  =&\frac{1}{4}(|0011\rangle|\varphi+\varphi'\rangle|\varphi+\varphi'\rangle+|1100\rangle|\varphi'-\varphi\rangle|\varphi'-\varphi\rangle
	  \\
	  &+|\varphi\rangle|\varphi\varphi\varphi-\varphi\varphi'\varphi'-\varphi'\varphi'\varphi+\varphi'\varphi\varphi'\rangle
	  \\
	  &+|\varphi'\rangle|\varphi\varphi'\varphi-\varphi\varphi\varphi'-\varphi'\varphi\varphi+\varphi'\varphi'\varphi'\rangle)_{1234E_{1}E_{2}E_{3}E_{4}}.\nonumber
\end{aligned}
\end{equation}

	The probability of getting the correct outcome (that is $|00\rangle$) by measuring via S on partite 1, 2 is $\frac{1}{4}$ and so the error rate is $\frac{3}{4}$. Similarly, the error rate for $b_{10p}$ is $\frac{3}{4}$.

\begin{equation}
\begin{aligned}
	b_{00}&=|00\rangle_{12}|00\rangle_{34}=|0000\rangle_{1234}=|0000\rangle_{1324}.
	\\
	 b_{00p}&=|00000000\rangle_{1324E_{1}E_{2}E_{3}E_{4}}=|00000000\rangle_{1234E_{1}E_{2}E_{3}E_{4}}.\nonumber
\end{aligned}
\end{equation}

    The probability of getting the correct outcome (that is $|00\rangle$) by measuring via S on partite 1, 2 is 1 and so the error rate is 0. Similarly, the error rate for $b_{11p}$ is 0.
	
	Now, the average error rate for Eve guess the order wrong is $\frac{1}{16}(4\times\frac{7}{10}+4\times\frac{1}{2}+4\times\frac{3}{4}+2\times\frac{1}{2}+2\times0)=\frac{93}{160}$ and so the whole error rate is $\frac{93}{320}$ which is larger than $\frac{1}{4}$, the error rate of attacking the BB84 protocol by purification.
	
	\subsection{Substituted attack}
	
	Let us focus on substituted attacks. Step 3 of protocol I is not needed in such case and so the efficiency or security can be increased. Eve might take a state sent by Alice herself, measuring it or keeping it until she obtains more information. However, Alice and Bob will not continue the procedure until Bob receives a state. Thus, Eve has to send another state to Bob, instead. Eve could get enough information after stealing the A partita in protocol I or after Alice publishes the order string s in protocol II. She can measure the state via basis S and know exactly what Alice sent. But Bob can detect her, for the reason that Eve might substitute a product state by an entangled state or substitute an entangled state by a product state. In both cases, Bob's measurement outcome could be incorrect.
	
	In more details, in protocol I, if Eve steals the B partita of the state and sends one partita of her state, instead. Since, at this time, Eve is not able to discriminate whether the state is separated or entangled, she can not send a state to Bob and transform it into the state she needs after stealing the A partita, since local transformations (local unitary operations) of a state can not generate or break entanglement. For example, assume that Eve sends a partita of a product state, says the partita $E_{1}$ of $|0\rangle_{E_{1}}|0\rangle_{E_{2}}$ to Bob and keeps the state sent by Alice in step 4. If after stealing the partita A of the state and finding that the state is entangled, for example, be $|\varphi\rangle$, she can not prevent her being detectable. For now, no matter what state she sends to Bob, Bob will get a product state and there is at least with probability $\frac{1}{2}$ he will obtain an error outcome when measuring via basis S. The same argument is suitable if Eve sends a maximally entangled state, instead. In such case, if she finds that Alice sent a product state, she can do nothing to decrease the error rate of Bob less than $\frac{1}{2}$. Hence, the error rate is not less than $\frac{1}{4}$, for Eve wrongly guesses the state sent by Alice is entangled or separated and then Bob's measurement outcome is incorrect, if so.
	
	These arguments are also held for protocol II. Since Eve can not know the order of the states before sending her states to Bob, she might send a product state instead of an entangled state or conversely. If so, the error rate of Bob's checking procedure will not less than $\frac{1}{2}$ and the whole error rate will not less than $\frac{1}{4}$.
	
	It is worth noting that protocol I without step 3 can not employ the four Bell states instead of S. The four Bell states can be transformed into each other via local transformations. For this reason, Eve can steal the partita B of states sent by Alice and send a partita of maximally entangled states of her in step 4. Then she can send the other partita with transformations depending on outcomes of her Bell measurements after stealing the partita A.
	
	\subsection{Measure-resend attack}
	
	The security analyses for such attacks are similar to above. If Eve measures via single qubits and resends the states to Bob, then Bob will receive product states. Thus, Bob can detect Eve by the outcomes of entangled states sent by Alice. The probability would be at least $\frac{1}{4}$ for half of checking states be entangled and with probability $\frac{1}{2}$ be incorrect for those states. To against such an attack, step 3 of protocol I is not needed, similar to above.
	
	Eve might choose to measure via two-qubit states. Such an attack can only happen in protocol II, since Eve can never obtain both partite of states in protocol I when employing a measure-resend attack only. In protocol II, since Eve could not know the orders of states before she finishes her measurement and resends states to Bob, she might change correlations of states. For example, let Eve measure states sent by Alice via basis S, the only basis for Eve might gather the secret key without increasing errors. If she guesses the order wrongly, she will measure partite 1, 3 and partite 2, 4, respectively. If partite 1, 2 are entangled, after Eve's interaction, they become separated. Hence, the probability for Bob obtains an error for such a state is at least $\frac{1}{2}$. If the partite are separated, without loss generality assume that the state is $|00\rangle$, then there are three cases including Alice sends $|00\rangle_{12}|00\rangle_{34}$, $|00\rangle_{12}|11\rangle_{34}$ or $|00\rangle_{12}|\varphi\rangle_{34}$. Eve will not provide any error in the first case while Eve provides an error with probability $\frac{1}{2}$ in the second and $\frac{3}{4}$ in the third. The arrange error rate is not less than $\frac{1}{4}$.
	
	\subsection{Two stages attack}
	
	Two stages attacks only suitable for protocol I, since in protocol II, Alice sends states to Bob in only one stage. Let us assume that Eve attacks protocol I with different strategies in step 4 and step 6, when Alice sends different partite to Bob. If Eve employs a substituted attack in step 4, she might substitute product states instead of entangled states or substitute entangled states instead of product states, which would increase error rates as in substituted attacks only. If Eve employs measure-resend attacks in step 4, she might break entanglement of states, and errors would be occurred as in measure-resend attacks only.
	
	The only case left is that Eve purifies states in step 4, and measures states in step 6. For example, Eve might purify states via basis $\left \{ |j\rangle |j=0,1 \right \}$ in step 4, twice. The states become
	\qquad $|0000\rangle_{ABEE'}$, \quad $|1111\rangle_{ABEE'}$,
	
	\noindent $|\varphi^{p}\rangle=\frac{1}{\sqrt{2}}|0111-1000\rangle_{ABEE'}=\frac{1}{\sqrt{2}}|\varphi\varphi'-\varphi'\varphi\rangle_{AEBE'}$,
	
	\noindent $|\varphi'^{p}\rangle=\frac{1}{\sqrt{2}}|0111+1000\rangle_{ABEE'}=\frac{1}{\sqrt{2}}|\varphi'\varphi'-\varphi\varphi\rangle_{AEBE'}$
	
	\noindent after Eve operates bit-flip gates on the partita E', respectively and corresponding to $|00\rangle_{AB}, |11\rangle_{AB}, |\varphi\rangle_{AB}, |\varphi'\rangle_{AB}$. After Eve steals the partita A, she measures the partita AE via basis S. If, for a state, the outcome is $|00\rangle$ or $|11\rangle$, she knows that the state is $|00\rangle$ or $|11\rangle$, respectively and she sends the partita E' to Bob. If the outcome is $|\varphi'\rangle$, she operates nothing and sends the partita E' to Bob while if the outcome is $|\varphi\rangle$, she operates the phase-flip gate on the partita E' and sends the partita E' to Bob. Hence, the attack is undetectable without step 3 and Eve obtains all information on product states. That is why Alice employs step 3, the decoy states. With these states, Eve's purified attack above can be detected. For such an attack, when Bob measures decoy states via basis $\left \{ |+\rangle, |-\rangle \right \}$, outcomes are incorrect with half of the probability. That is the error rate of decoy states is $\frac{1}{2}$, which can be employed to detect Eve.
	
	\section{Implementing protocols over noisy channels}
	
	Let us discuss implementing the protocols over noisy channels, including collective dephasing(CD) noises, collective rotation (CR) noises, Pauli noises, amplitude damping(AD) and phase damping(PD) noises as well as mixtures of them. Instead of discussing fidelities or estimating noises of the channels, we modify the protocols such that they could be implemented in noisy channels as in noiseless ones without errors. The discussions follow the testing state method in \cite{SZ2021Entanglement}.
	
	\subsection{Collective dephasing (CD)}
	Collective dephasing noises assume that the whole protocol is implemented in a same-time cycle and so the noise affects each qubit equivalently via
	$\begin{bmatrix}
		1  & 0\\
		0  & e^{i\phi}
	\end{bmatrix}$
	under the computational basis, where $\phi$ is the parameter depending on the noise. Previous results for such noises include \cite{LZ2009Fault,LD2008Efficient,SD2010Efficient,BG2004Robust,SZ2021Entanglement}.
	
	For such noises, we shall modify protocol I, step 3, substituting decoy states $|+\rangle$ by $|\varphi\rangle$, and protocol I step 7, measuring via S on decoy states, instead. Protocol II does not need to be modified. Now, Protocol I and protocol II completely immune such noises, since all states in the protocols are changed nothing but global phases when affecting by the noises. The consumption of states in protocol I now becomes 2.5N qubits for a N-bit key string while the consumption of states in protocol II remains unchanged.
	
	\subsection{Collective rotation (CR)}
	Collective rotation noises assume that the whole protocol is implemented in a same-time cycle and so the noise affects each qubit equivalently via
	$\begin{bmatrix}
		cos\theta  & sin\theta\\
		sin\theta  & -cos\theta
	\end{bmatrix}$
	under the computational basis, where $\theta$ is the parameter depending on the noise and evolving upon time. Previous results for such noises include \cite{LZ2009Fault,LD2008Efficient,SD2010Efficient,BG2004Robust,SZ2021Entanglement}.
	
	We shall modify protocol I and II both. Firstly, we substitute S by $S'=\left\{ |\varphi\rangle_{AB}, |\varphi''\rangle_{AB}=\frac{1}{\sqrt{2}}|00+11\rangle_{AB} \right\}$ in both protocol I and protocol II. Then in protocol I, decoy states in step 3 are substituted by $|\varphi\rangle_{B}$. And in step 7, Bob measures via S', instead. The reason for employing S' instead of S is that states in S' are are not changed when affecting by CR while states in S might not. The consumption of states now becomes 5N qubits for protocol I and 4N qubits for protocol II for a N-bit key string.
	
	\subsection{Pauli noises}
	Pauli noises act on each qubit via Pauli operators,
	I=$\begin{bmatrix}
		1  & 0\\
		0  & 1
	\end{bmatrix}$,
	Z=$\begin{bmatrix}
		1  & 0\\
		0  & -1
	\end{bmatrix}$,
	X=$\begin{bmatrix}
		0  & 1\\
		1  & 0
	\end{bmatrix}$,
	ZX=$\begin{bmatrix}
		0  & 1\\
		-1  & 0
	\end{bmatrix}$.
	under the computational basis, with probabilities $p_{I}$, $p_{Z}$, $p_{X}$, $p_{ZX}$, summing to 1. Previous works include \cite{SS2007Degenerate,FW2008Lower,FM2001Enhanced,CR2011Experimental,SZ2021Entanglement}.
	
	\subsubsection{One Pauli channel}
	In one Pauli channels, states suffer two of the four Pauli operators with one of which is I.
	
	Let us assume that states suffer I with probability p and Z with probability 1-p.
	
	As for protocol I, to deal with the noise, Alice prefers to send states assisted with auxiliary partite. She sends two partite together and assumes that they suffer the same effects. In more details, Alice employs states
	
	\quad $|00++\rangle_{ABA'B'}$, \qquad $|11++\rangle_{ABA'B'}$,
	
    \quad $|\varphi++\rangle_{ABA'B'}=\frac{1}{\sqrt{2}}|(01-10)++\rangle_{ABA'B'}$,
	
    \quad $|\varphi'++\rangle_{ABA'B'}=\frac{1}{\sqrt{2}}|(01+10)++\rangle_{ABA'B'}$,
	
	 \noindent
	 instead of $|00\rangle_{AB}, |11\rangle_{AB}$, $|\varphi\rangle_{AB}$, $|\varphi'\rangle_{AB}$, respectively. Alice always sends partite A, A' together and B, B' together and assumes that partite A, A' always suffer the same effects while partite B, B' always suffer the same effects. The same method and assumption are also applied for decoy states in step 3, that is, Alice employs $|++\rangle_{BB'}$ as decoy states. After Bob receives states (step 4 and step 6), he measures partite A', and B', respectively, via basis $\left\{|+\rangle, |-\rangle \right\}$. If the outcomes are $|+\rangle$, he does nothing, while he transforms the partita A or B via the Pauli operator Z if the outcome is $|-\rangle$ on the partita A' or B', respectively, and then continues the procedure (step 7). Hence, the auxiliary partite are employed for detecting whether states are influenced by Z and if so, Bob corrects them via Z.
	
	As for protocol II, Alice does nothing but always sends a block of four qubits together and assumes that they suffer the same effects. Now all states are changed nothing but global phases, which affect nothing.
	
	These arguments are also suitable for states suffering I with probability p and X with probability 1-p. Alice and Bob only need to employ $|+\rangle$, $|-\rangle$ instead of $|0\rangle$, $|1\rangle$, that is Alice provides H=$\begin{bmatrix}
		1  & 1\\
		1  & -1
	\end{bmatrix}$
	on all partite of states she creates and Bob operates H on all partite he obtains before measuring. And for states suffering I with probability p and ZX with probability 1-p, they provide operator
	H'=$\begin{bmatrix}
		1  & 1\\
		i  & -i
	\end{bmatrix}$
	instead of H.
	
	\subsubsection{Two Pauli channel}

    In two Pauli channels, $p_{I}$, $p_{Z}$, $p_{X}$, $p_{ZX}$ might all be non-zero, and so the final states might be completely mixed. We follow the above method to deal with such noises.

    For protocol I, Alice employs states

    \noindent $|00++00\rangle_{ABA'B'A''B''}$, \qquad $|11++00\rangle_{ABA'B'A''B''}$,

    \noindent $|\varphi++00\rangle_{ABA'B'A''B''}=\frac{1}{\sqrt{2}}|(01-10)++00\rangle_{ABA'B'A''B''}$,

    \noindent$|\varphi'++00\rangle_{ABA'B'A''B''}=\frac{1}{\sqrt{2}}|(01+10)++00\rangle_{ABA'B'A''B''}$
    \noindent instead of $|00\rangle_{AB}, |11\rangle_{AB},\\ |\varphi\rangle_{AB}, |\varphi'\rangle_{AB}$, respectively, and decoy states become $|++0\rangle_{BB'B''}$, instead of $|+\rangle_{B}$.
    Alice always sends partite A, A', A'' together and B, B', B'' together to assume that partite A, A', A'' always suffer the same effects while partite B, B', B'' always suffer the same effects. After Bob receives states (step 4 and step 6), he measures partite A', A'' and B', B'', respectively, via basis $\left\{|+\rangle, |-\rangle \right\}$ for A', B' and basis $\left\{|0\rangle, |1\rangle \right\}$ for A'', B''. If the outcomes on A', A'' are $|+\rangle$ and $|0\rangle$, he does nothing, while he transforms the partita A via the operator Z, X, or XZ if the outcomes are $|-\rangle$ and $|0\rangle$, $|+\rangle$ and $|1\rangle$, or $|-\rangle$ and $|1\rangle$ on A' and A'', respectively. The same manipulations are also applied for partite B, B' and B''. Then he continues the procedure (step 7). Hence, the auxiliary partite are employed for detecting whether states are influenced by Z, X, ZX and if so, Bob corrects them via Z, X, XZ, respectively.

	As for protocol II, Alice only needs to employ exactly one auxiliary partita for each block while qubits of the auxiliary partita are always $|0\rangle$. Bob operates X on the corresponding ordinary state when getting the outcome $|1\rangle$ by measuring via basis $\left\{|0\rangle, |1\rangle \right\}$ on the auxiliary partita of a state while he does nothing, otherwise. This is because that operator Z affects nothing for protocol II and so they only need to confirm whether operator X is applied, that is whether states are affected by X or ZX. Other procedures are similar to above.
	
	\subsection{Phase damping (PD) and Amplitude damping (AD)}
	
	Phase damping noises have kraus operators
	$E_{0}=\begin{bmatrix}
		\sqrt{1-p}  & 0\\
		0           & \sqrt{1-p}
	\end{bmatrix}$
	$E_{1}=\begin{bmatrix}
		\sqrt{p}  & 0\\
		0  & 0
	\end{bmatrix}$
	$E_{2}=\begin{bmatrix}
		0  & 0\\
		0  & \sqrt{p}
	\end{bmatrix}$. Hence, a state sent by the channel has probability 1-p remains unchanged and probability p suffers errors.
	
	Amplitude damping noises have kraus operators
	$E_{0}=\begin{bmatrix}
		1  & 0\\
		0  & \sqrt{1-p}
	\end{bmatrix}$
	$E_{1}=\begin{bmatrix}
		0  & \sqrt{p}\\
		0  & 0
	\end{bmatrix}$.
    Hence, a state sent by the channel has probability 1-p suffers $E_{0}$ and probability p suffers $E_{1}$. Previous works include \cite{TP2015Applications,SS2015Controlled,OS2013Dissipative,TM2000Decoherence,XY2016Protecting,SZ2021Entanglement}.
	
	For such noises, Alice can employ states
	
	\qquad $|0101\rangle_{AA'BB'}$, \qquad $|1010\rangle_{AA'BB'}$,
	
	\qquad $|\psi\rangle_{AA'BB'}=\frac{1}{\sqrt{2}}|0110-1001\rangle_{AA'BB'}$,
	
	\qquad $|\psi'\rangle_{AA'BB'}=\frac{1}{\sqrt{2}}|0110+1001\rangle_{AA'BB'}$
	
	\noindent instead of $|00\rangle_{AB}, |11\rangle_{AB}, |\varphi\rangle_{AB}, |\varphi'\rangle_{AB}$, respectively, and decoy states $|\varphi'\rangle_{BB'}$ instead of $|+\rangle_{B}$. That is, Alice employs an auxiliary partita for each partita, setting ordinary states of auxiliary partite be $|1\rangle$, then provides C-NOT gates on auxiliary partite. Alice always sends partite A and A', partite B and B' together to assume that partite A and A' always suffer the same effects of noises and so do partite B and B'. For Bob, after receiving states, he firstly discards states with partite lost. The states left are those not affected by noises. Bob then provides C-NOT gates on those states and continuous step 7. The above arguments are suitable for both protocol I and protocol II.
	
	\subsection{Mixture of noises}
	
	To deal with mixtures of noises, one might employ a combination of the above strategies\cite{SZ2021Entanglement} or consider a technique called decoherence-free subspace\cite{BE2004Decoherence,WA2003Decoherence,YZ2008Decoy,ZY2006Experimental,W2003On,ZR1997Noiseless,KY2013Robustness,C2007Six}. Decoherence-free states are invariant under collective noises and our arguments for AD and PD noises shall also be applied. However, there is a unique decoherence-free state in a two-qubit system which can not be employed for coding. For a four-qubit system, $C^{4}\otimes C^{4}$, decoherence-free subspace is of 2-dimensional. Thus, coding might be applied in such a system with orthogonal states.
	
	\section{Conclusion}
	
	In this paper, we presented two quantum key distribution protocols based on orthogonal state encoding. Protocol I consumes more but does not need to employ an order-rearrangement technique while protocol II consumes less with an order-rearrangement technique. Both protocols employ the same set of coding states, of which half are maximally entangled and the other half are separated. We demonstrated the advantages on bit consumptions of the protocols and compare them with certain previous ones. We also provided security analyses for them under some special attacks and modified them for implementing over noisy channels.
	
	Our protocols, on one hand, are highly efficient on the consumption of both qubits and classical bits. In protocol II, all states are employed for key states except those for checking, which are always assumed to be half of the states like in the BB84 protocol, while in protocol I, another 12.5 percents of states are employed as decoy states but it does not need to employ order-rearrangement techniques. Our protocols does not need to employ a full Bell measurement which might be lowly efficient, since the employed states are half maximally entangled and half separated. On the other hand, our protocols can be modified over noisy environments by applying the testing state method.

     Finally, note that a main obstacle in implementing such protocols, including the two protocols, other order-rearrangement ones or even the BB84 protocol with a Haramard gate, is the need of a quantum memory. Therefore, investigations on the quantum memories might be significant. On the other hand, researches in other noises, even for the general noise, as well as experimental realizations of the protocols are considerable.
	  		
	\section*{Data availability}
	The author declare that all data supporting the findings of this study are available within the paper.
	
	\section*{Competing interests}
	\qquad The author declare no competing interests.
	
	\section*{Funding}
	\qquad No funding.
	
	\bibliographystyle{unsrt}
    \bibliography{Bibliog}

\begin{thebibliography}{10}

\bibitem{S1994Algorithms}
P.~Shor.
\newblock Algorithms for quantum computation: discrete logarithms and
  factoring.
\newblock {\em In Proceedings of 35th Annual Symposium on the Foundations of
  Computer Science, IEEE Computer Society Press, Los Alamitos, CA}, pages
  124--134, 1994.

\bibitem{BB1984Quantum}
C.~H. Bennett and G.~Brassard.
\newblock Quantum cryptography: Public key distribution and coin tossing.
\newblock In {\em In Proceedings of IEEE International Conference on
  Computers}, 1984.

\bibitem{SP2000Simple}
P.~W. Shor and J.~Preskill.
\newblock Simple proof of security of the bb84 quantum key distribution
  protocol.
\newblock {\em Physical Review Letters}, 85(2):441--444, 2000.

\bibitem{E1991Quantum}
A.~K. Ekert.
\newblock Quantum cryptography based on bell's theorem.
\newblock {\em Physical Review Letters}, 67(6):661, 1991.

\bibitem{BB1992Quantum}
C.~Bennett, G.~Brassard, and N.~Mermin.
\newblock Quantum cryptography without bell's theorem.
\newblock {\em In Proceedings of 35th Annual Symposium on the Foundations of
  Computer Science, IEEE Computer Society Press, Los Alamitos, CA}, pages
  124--134, 1994.

\bibitem{B1998Optimal}
D.~Bruss.
\newblock Optimal eavesdropping in quantum cryptography with six states.
\newblock {\em Physical Review Letters}, 81:3018, 1998.

\bibitem{CB2002Security}
N.~J. Cerf, M.~Bourennane, A.~Karlsson, and N.~Gisin.
\newblock Security of quantum key distribution using d-level systems.
\newblock {\em Physical Review Letters}, 88(12):127902, 2002.

\bibitem{LZ2022Toward}
W.~Z. Liu, Y.~Z. Zhang, Y.~Z. Zhen, M.~H. Li, Y.~Liu, J.~Y. Fan, F.~H. Xu,
  Q.~Zhang, and J.~W. Pan.
\newblock Toward a photonic demonstration of device-independent quantum key
  distribution.
\newblock {\em Physical Review Letter}, 129:050502, Jul 2022.

\bibitem{ZL2022A}
W.~Zhang, T.~V. Leent, K.~Redeker, R.~Garthoff, R.~Schwonnek, F.~Fertig,
  S.~Eppelt, W.~Rosenfeld, V.~Scarani, C.~W. Lim, and Weinfurter.
\newblock A device-independent quantum key distribution system for distant
  users.
\newblock {\em Nature}, 607:687, 2022.

\bibitem{XL2021Overcoming}
Y.~M. Xie, B.~H. Li, Y.~S. Lu, X.~Y. Cao, W.~B. Liu, H.~L. Yin, and Z.~B. Chen.
\newblock Overcoming the rate--distance limit of device-independent quantum key
  distribution.
\newblock {\em Opt. Lett.}, 46(7):1632--1635, Apr 2021.

\bibitem{YC2016Measurement}
H.~L. Yin, T.~Y. Chen, Z.~W. Yu, H.~Liu, L.~X. You, Y.~H. Zhou, S.~J. Chen,
  Y.~Q. Mao, M.~Q. Huang, W.~J. Zhang, H.~Chen, M.~J. Li, D.~Nolan, F.~Zhou,
  X.~Jiang, Z.~Wang, Q.~Zhang, X.~B. Wang, and J.~W. Pan.
\newblock Measurement-device-independent quantum key distribution over a 404 km
  optical fiber.
\newblock {\em Physical Review Letter}, 117:190501, Nov 2016.

\bibitem{FL2021Measurement}
G.~J. Fan-Yuan, F.~Y. Lu, S.~Wang, Z.~Q. Yin, D.~Y. He, Z.~Zhou, J.~Teng,
  W.~Chen, G.~C. Guo, and Z.~F. Han.
\newblock Measurement-device-independent quantum key distribution for
  nonstandalone networks.
\newblock {\em Photon. Res.}, 9(10):1881--1891, Oct 2021.

\bibitem{XL2022Breaking}
Y.~M. Xie, Y.~S. Lu, C.~X. Weng, X.~Y. Cao, Z.~Y. Jia, Y.~Bao, Y.~Wang, Y.~Fu,
  H.~L. Yin, and Z.~B. Chen.
\newblock Breaking the rate-loss bound of quantum key distribution with
  asynchronous two-photon interference.
\newblock {\em PRX Quantum}, 3:020315, Apr 2022.

\bibitem{LY2018Overcoming}
M.~Lucamarini, Z.~L. Yuan, J.~F. Dynes, and A.~J. Shields.
\newblock Overcoming the rate–distance limit of quantum key distribution
  without quantum repeaters.
\newblock {\em Nature}, 557(7705):400–403, May 2018.

\bibitem{WY2022Twin}
S.~Wang, Z.~Q. Yin, D.~Y. He, W.~Chen, R.~Q. Wang, P.~Ye, Y.~Zhou, G.~J.
  Fan-Yuan, F.~X. Wang, W.~Chen, Y.~G. Zhu, P.~V. Morozov, A.~V. Divochiy,
  Z.~Zhou, G.~C. Guo, and Z.~F. Han.
\newblock Twin-field quantum key distribution over 830-km fibre.
\newblock {\em Nature Photonics}, 16:154, 2022.

\bibitem{LL2021Homodyne}
W.~B. Liu, C.~L. Li, Y.~M. Xie, C.~X. Weng, J.~Gu, X.~Y. Cao, Y.~S. Lu, B.~H.
  Li, H.~L. Yin, and Z.~B. Chen.
\newblock Homodyne detection quadrature phase shift keying continuous-variable
  quantum key distribution with high excess noise tolerance.
\newblock {\em PRX Quantum}, 2:040334, Nov 2021.

\bibitem{LZ2022Automated}
Z.~P. Liu, M.~G. Zhou, W.~B. Liu, C.~L. Li, J.~Gu, H.~L. Yin, and Z.~B. Chen.
\newblock Automated machine learning for secure key rate in discrete-modulated
  continuous-variable quantum key distribution.
\newblock {\em Opt. Express}, 30(9):15024--15036, Apr 2022.

\bibitem{WB2014A}
N.~Walenta, A.~Burg, D.~Caselunghe, J.~Constantin, N.~Gisin, O.~Guinnard,
  R.~Houlmann, P.~Junod, B.~Korzh, N.~Kulesza, M.~Legré, C.~W. Lim, T.~Lunghi,
  L.~Monat, C.~Portmann, M.~Soucarros, R.~T. Thew, P.~Trinkler, G.~Trolliet,
  F.~Vannel, and H.~Zbinden.
\newblock A fast and versatile quantum key distribution system with hardware
  key distillation and wavelength multiplexing.
\newblock {\em New Journal of Physics}, 16(1):013047, jan 2014.

\bibitem{GX2022Simple}
R.~Q. Gao, Y.~M. Xie, J.~Gu, W.~B. Liu, C.~X. Weng, B.~H. Li, H.~L. Yin, and
  Z.~B. Chen.
\newblock Simple security proof of coherent-one-way quantum key distribution.
\newblock {\em Opt. Express}, 30(13):23783--23795, Jun 2022.

\bibitem{FL2022Robust}
G.~J. Fan-Yuan, F.~Y. Lu, S.~Wang, Z.~Q. Yin, D.~Y. He, W.~Chen, Z.~Zhou, Z.~H.
  Wang, J.~Teng, G.~C. Guo, and Z.~F. Han.
\newblock Robust and adaptable quantum key distribution network without trusted
  nodes.
\newblock {\em Optica}, 9(7):812--823, Jul 2022.

\bibitem{LL2002Theoretically}
G.~L. Long and X.~S. Liu.
\newblock Theoretically efficient high-capacity quantum-key-distribution
  scheme.
\newblock {\em Physical Review A}, 65:032302, Feb 2002.

\bibitem{ZD2017Quantum}
W.~Zhang, D.~S. Ding, Y.~B. Sheng, L.~Zhou, B.~S. Shi, and G.~C. Guo.
\newblock Quantum secure direct communication with quantum memory.
\newblock {\em Physical Review Letter}, 118:220501, May 2017.

\bibitem{DL2003Two}
F.~G. Deng, G.~L. Long, and X.~S. Liu.
\newblock Two-step quantum direct communication protocol using the
  einstein-podolsky-rosen pair block.
\newblock {\em Physical Review A}, 68:042317, Oct 2003.

\bibitem{BF2002Deterministic}
Kim Bostr\"om and Timo Felbinger.
\newblock Deterministic secure direct communication using entanglement.
\newblock {\em Physical Review Letter}, 89:187902, Oct 2002.

\bibitem{BE2001Secure}
A.~Beige, B.~G. Englert, C.~Kurtsiefer, and H.~Weinfurter.
\newblock Secure communication with a publicly known key.
\newblock {\em Acta Physica Polonica A}, 101:357--368, 2002.

\bibitem{WL2022Quantum}
J.~W. Wu, G.~L. Long, and M.~Hayashi.
\newblock Quantum secure direct communication with private dense coding using a
  general preshared quantum state.
\newblock {\em Physical Review Applied}, 17(6):064011, jun 2022.

\bibitem{PL2020Experimental}
D.~Pan, Z.~S. Lin, J.~W. Wu, H.~R. Zhang, Z.~Sun, D.~Ruan, L.~G. Yin, and G.~L.
  Long.
\newblock Experimental free-space quantum secure direct communication and its
  security analysis.
\newblock {\em Photonics Research}, 8(9):1522, aug 2020.

\bibitem{WZ2006Quantum}
J.~Wang, Q.~Zhang, and C.~J. Tang.
\newblock Quantum secure direct communication based on order rearrangement of
  single photons.
\newblock {\em Physics Letters A}, 358(4):256--258, oct 2006.

\bibitem{DL2007Quantum}
F.~G. Deng, X.~H. Li, C.~Y. Li, P.~Zhou, and H.~Y. Zhou.
\newblock Quantum secure direct communication network with superdense coding
  and decoy photons.
\newblock {\em Physica Scripta}, 76(1):25--30, jun 2007.

\bibitem{BB1993Teleporting}
C.~H. Bennett, G.~Brassard, C.~Crépeau, R.~Jozsa, A.~Peres, and W.~K.
  Wootters.
\newblock Teleporting an unknown quantum state via dual classical and
  einstein-podolsky-rosen channels.
\newblock {\em Physical Review Letters}, 70(13):1895--1899, 1993.

\bibitem{GV1995Quantum}
L.~Goldenberg and L.~Vaidman.
\newblock Quantum cryptography based on orthogonal states.
\newblock {\em Physical Review Letters}, 75(7):1239–1243, Aug 1995.

\bibitem{DL2003Controlled}
F.~G. Deng and G.~L. Long.
\newblock Controlled order rearrangement encryption for quantum key
  distribution.
\newblock {\em Physical Review A}, 68(4):042315, Oct 2003.

\bibitem{SA2014Protocols}
C.~Shukla, N.~Alam, and A.~Pathak.
\newblock Protocols of quantum key agreement solely using bell states and bell
  measurement.
\newblock {\em Quantum Information Processing}, 13(11):2391–2405, Jul 2014.

\bibitem{ZX2006Secure}
A.~D. Zhu, Y.~Xia, Q.~B. Fan, and S.~Zhang.
\newblock Secure direct communication based on secret transmitting order of
  particles.
\newblock {\em Physical Review A}, 73(2):457--460, 2006.

\bibitem{YS2014Two}
P.~Yadav, R.~Srikanth, and A.~Pathak.
\newblock Two-step orthogonal-state-based protocol of quantum secure direct
  communication with the help of order-rearrangement technique.
\newblock {\em Quantum Information Processing}, 13(12):2731–2743, Sep 2014.

\bibitem{SP2013Beyond}
C.~Shukla, A.~Pathak, and R.~Srikanth.
\newblock Beyond the goldenberg-vaidman protocol: Secure and efficient quantum
  communication using arbitrary, orthogonal, multi-particle quantum states.
\newblock {\em International Journal of Quantum Information},
  10(08):1241009--1241009--13, 2013.

\bibitem{GL2001Quantum}
G.~P. Guo, C.~F. Li, B.~S. Shi, J.~Li, and G.~C. Guo.
\newblock Quantum key distribution scheme with orthogonal product states.
\newblock {\em Physical Review A}, 64(4):042301, Sep 2001.

\bibitem{AB2014Orthogonal}
S.~Aravinda, A.~Banerjee, A.~Pathak, and R.~Srikanth.
\newblock Orthogonal-state-based cryptography in quantum mechanics and local
  post-quantum theories.
\newblock {\em International Journal of Quantum Information},
  12(07n08):1560020, Nov 2014.

\bibitem{H2011Quantum}
G.~P. He.
\newblock Quantum key distribution based on orthogonal states allows secure
  quantum bit commitment.
\newblock {\em Journal of Physics A: Mathematical and Theoretical},
  44(44):445305, Oct 2011.

\bibitem{SB2016Secure}
C.~Shukla, A.~Banerjee, A.~Pathak, and R.~Srikanth.
\newblock Secure quantum communication with orthogonal states.
\newblock {\em International Journal of Quantum Information}, 14(06):1640021,
  2016.

\bibitem{N2009Counterfactual}
T.~G. Noh.
\newblock Counterfactual quantum cryptography.
\newblock {\em Physical Review Letters}, 103(23):230501, 2009.

\bibitem{AB2010Experimental}
A.~Avella, G.~Brida, I.~P. Degiovanni, M.~Genovese, M.~Gramegna, and P.~Traina.
\newblock Experimental quantum-cryptography scheme based on orthogonal states.
\newblock {\em Physical Review A}, 82(6):062309, Dec 2010.

\bibitem{AS2013Semi}
A.~Shenoy, R.~Srikanth, and T.~Srinivas.
\newblock Semi-counterfactual cryptography.
\newblock {\em Europhysics Letters}, 103(6):60008, Sep 2013.

\bibitem{LZ2009Fault}
X.~H. Li, B.~K. Zhao, Y.~B. Sheng, F.~G. Deng, and H.~Y. Zhou.
\newblock Fault tolerant quantum key distribution based on quantum dense coding
  with collective noise.
\newblock {\em International Journal of Quantum Information}, 7(08):1479--1489,
  2009.

\bibitem{LD2008Efficient}
X.~H. Li, F.~G. Deng, and H.~Y. Zhou.
\newblock Efficient quantum key distribution over a collective noise channel.
\newblock {\em Physical Review A}, 78(2):022321, Aug 2008.

\bibitem{SD2010Efficient}
Y.~B. Sheng and F.~G. Deng.
\newblock Efficient quantum entanglement distribution over an arbitrary
  collective-noise channel.
\newblock {\em Physical Review A}, 81(4):042332, Apr 2010.

\bibitem{BG2004Robust}
J.~C. Boileau, D.~Gottesman, R.~Laflamme, D.~Poulin, and R.~W. Spekkens.
\newblock Robust polarization-based quantum key distribution over a
  collective-noise channel.
\newblock {\em Physical Review Letters}, 92(1):017901, Jan 2004.

\bibitem{SS2007Degenerate}
G.~Smith and J.~A. Smolin.
\newblock Degenerate quantum codes for pauli channels.
\newblock {\em Physical Review Letters}, 98(3):030501, Jan 2007.

\bibitem{FW2008Lower}
J.~Fern and K.~B. Whaley.
\newblock Lower bounds on the nonzero capacity of pauli channels.
\newblock {\em Physical Review A}, 78(6), Dec 2008.

\bibitem{FM2001Enhanced}
D.~G. Fischer, H.~Mack, M.~A. Cirone, and M.~Freyberger.
\newblock Enhanced estimation of a noisy quantum channel using entanglement.
\newblock {\em Physical Review A}, 64(2):022309, Jul 2001.

\bibitem{CR2011Experimental}
A.~Chiuri, V.~Rosati, G.~Vallone, S.~Pádua, H.~Imai, S.~Giacomini,
  C.~Macchiavello, and P.~Mataloni.
\newblock Experimental realization of optimal noise estimation for a general
  pauli channel.
\newblock {\em Physical Review Letters}, 107(25):253602, Dec 2011.

\bibitem{TP2015Applications}
K.~Thapliyal and A.~Pathak.
\newblock Applications of quantum cryptographic switch: various tasks related
  to controlled quantum communication can be performed using bell states and
  permutation of particles.
\newblock {\em Quantum Information Processing}, 14(7):2599–2616, Apr 2015.

\bibitem{SS2015Controlled}
V.~Sharma, C.~Shukla, S.~Banerjee, and A.~Pathak.
\newblock Controlled bidirectional remote state preparation in noisy
  environment: a generalized view.
\newblock {\em Quantum Information Processing}, 14(9):3441–3464, Jun 2015.

\bibitem{OS2013Dissipative}
S.~Omkar, R.~Srikanth, and S.~Banerjee.
\newblock Dissipative and non-dissipative single-qubit channels: dynamics and
  geometry.
\newblock {\em Quantum Information Processing}, 12(12):3725–3744, Aug 2013.

\bibitem{TM2000Decoherence}
Q.~A. Turchette, C.~J. Myatt, B.~E. King, C.~A. Sackett, D.~Kielpinski, W.~M.
  Itano, C.~Monroe, and D.~J. Wineland.
\newblock Decoherence and decay of motional quantum states of a trapped atom
  coupled to engineered reservoirs.
\newblock {\em Physical Review A}, 62(5):53807.

\bibitem{XY2016Protecting}
X.~Xiao, Y.~Yao, Y.~M. Xie, X.~H. Wang, and Y.~L. Li.
\newblock Protecting entanglement from correlated amplitude damping channel
  using weak measurement and quantum measurement reversal.
\newblock {\em Quantum Information Processing}, 15(9):3881--3891, 2016.

\bibitem{ST2015Which}
R.~D. Sharma, K.~Thapliyal, A.~Pathak, A.~K. Pan, and A.~De.
\newblock Which verification qubits perform best for secure communication in
  noisy channel?
\newblock {\em Quantum Information Processing}, 15(4):1703–1718, Dec 2015.

\bibitem{SB2008Squeezed}
R.~Srikanth and S.~Banerjee.
\newblock Squeezed generalized amplitude damping channel.
\newblock {\em Physical Review A}, 77(1):012318, Jan 2008.

\bibitem{SO2012The}
N.~Srinatha, S.~Omkar, R.~Srikanth, S.~Banerjee, and A.~Pathak.
\newblock The quantum cryptographic switch.
\newblock {\em Quantum Information Processing}, 13(1):59–70, Sep 2012.

\bibitem{TB2015Quasiprobability}
K.~Thapliyal, S.~Banerjee, A.~Pathak, S.~Omkar, and V.~Ravishankar.
\newblock Quasiprobability distributions in open quantum systems: Spin-qubit
  systems.
\newblock {\em Annals of Physics}, 362:261–286, Nov 2015.

\bibitem{TB2016Tomograms}
K.~Thapliyal, S.~Banerjee, and A.~Pathak.
\newblock Tomograms for open quantum systems: In(finite) dimensional optical
  and spin systems.
\newblock {\em Annals of Physics}, 366:148–167, Mar 2016.

\bibitem{SZ2021Entanglement}
H.~Shu, C.~Y. Zhang, Y.~Q. Chen, and Z.~J. Zheng.
\newblock Entanglement-based quantum key distribution over noisy channels.
\newblock 2021.

\bibitem{C2000Quantum}
A.~Cabello.
\newblock Quantum key distribution in the holevo limit.
\newblock {\em Physical review letters}, 85:5635--5638, Dec 2000.

\bibitem{BE2004Decoherence}
M.~Bourennane, M.~Eibl, S.~Gaertner, C.~Kurtsiefer, A.~Cabello, and
  H.~Weinfurter.
\newblock Decoherence-free quantum information processing with four-photon
  entangled states.
\newblock {\em Physical Review Letters}, 92(10), Mar 2004.

\bibitem{WA2003Decoherence}
Z.~D. Walton, A.~F. Abouraddy, A.~V. Sergienko, B.~E.~A. Saleh, and M.~C.
  Teich.
\newblock Decoherence-free subspaces in quantum key distribution.
\newblock {\em Physical Review Letters}, 91(8):087901, Aug 2003.

\bibitem{YZ2008Decoy}
Z.~Q. Yin, Y.~B. Zhao, Z.~W. Zhou, Z.~F. Han, and G.~C. Guo.
\newblock Decoy states for quantum key distribution based on decoherence-free
  subspaces.
\newblock {\em Physical Review A}, 77(6):062326, Jun 2008.

\bibitem{ZY2006Experimental}
Q.~Zhang, J.~Yin, T.~Y. Chen, S.~Lu, J.~Zhang, X.~Q. Li, T.~Yang, X.~B. Wang,
  and J.~W. Pan.
\newblock Experimental fault-tolerant quantum cryptography in a
  decoherence-free subspace.
\newblock {\em Physical Review A}, 73(2):020301, Feb 2006.

\bibitem{W2003On}
X.~B. Wang.
\newblock On quantum key distribution in decoherence-free subspace.
\newblock {\em arXiv}, page 0308092, 2003.

\bibitem{ZR1997Noiseless}
P.~Zanardi and M.~Rasetti.
\newblock Noiseless quantum codes.
\newblock {\em Physical Review Letters}, 79(17):3306–3309, Oct 1997.

\bibitem{KY2013Robustness}
H.~Kumagai, T.~Yamamoto, M.~Koashi, and N.~Imoto.
\newblock Robustness of quantum communication based on a decoherence-free
  subspace using a counter-propagating weak coherent light pulse.
\newblock {\em Physical Review A}, 87(5):052325, May 2013.

\bibitem{C2007Six}
A.~Cabello.
\newblock Six-qubit permutation-based decoherence-free orthogonal basis.
\newblock {\em Physical Review A}, 75(2):020301, Feb 2007.

\end{thebibliography}

	\end{document}